\documentclass[]{svjour2}

\smartqed

\usepackage{amssymb}
\usepackage{graphicx}

\journalname{J Stat Phys}

\begin{document}

\title{Counter-Ions Near a Charged Wall: Exact Results for Disc and 
Planar Geometries}

\titlerunning{Counter-Ions Near a Charged Wall}

\author{Ladislav \v{S}amaj}

\institute{Institute of Physics, Slovak Academy of Sciences, 
D\'ubravsk\'a cesta 9, SK-84511 Bratislava, Slovakia \\
\email{Ladislav.Samaj@savba.sk}}

\date{Received:  / Accepted: }

\maketitle

\begin{abstract}
Macromolecules, when immersed in a polar solvent like water, become  
charged by a fixed surface charge density which is compensated by 
``counter-ions'' moving out of the surface. 
Such classical particle systems exhibit poor screening properties at any
temperature and the trivial bulk regime (far away from the charged surface)  
with no particles, so the validity of standard Coulomb sum rules 
is questionable.
In the present paper, we concentrate on the two-dimensional version of the
model with the logarithmic interaction potential. 
We go from the finite disc to the semi-infinite planar geometry. 
The system is exactly solvable for two values of the coupling constant 
$\Gamma$: in the Poisson-Boltzmann mean-field limit $\Gamma\to 0$ and at 
the free-fermion point $\Gamma=2$. 
We show that the finite-size expansion of the free energy does not contain 
universal term as is usual for Coulomb fluids.
For the coupling constant being an arbitrary positive even integer, 
using an anticommuting representation of the partition function and 
many-body densities we derive a sequence of sum rules. 
As a result, the contact density of counter-ions at the wall is available
for the disc.
The amplitude function, which characterizes the asymptotic inverse-power 
law behavior of the two-body density along the wall, is found to be related 
to the particle density profile.
The dielectric susceptibility tensor, calculated exactly for an arbitrary
coupling and the particle number, exhibits the anticipated disc value 
in the thermodynamic limit, in spite of zero contribution from the bulk region.
Some of the results obtained in the Poisson-Boltzmann limit are generalized 
to an arbitrary Euclidean dimension.


\end{abstract}

\renewcommand{\theequation}{1.\arabic{equation}}
\setcounter{equation}{0}

\section{Introduction} \label{Sect.1}
Experiments with macromolecules are usually performed in polar solvents
such as water (which is also the relevant medium for biological systems). 
The surface of the macromolecule releases micro-ions into the polar
solvent and in this way acquires a fixed surface charge density, opposite
to the charge of micro-ions (coined as ``counter-ions'').
For large macromolecules, the total surface charge can be of order
of thousands elementary charges $e$; in the first approximation, the curved
surface of the macromolecule can be replaced by an infinite rectilinear plane.
The charged macromolecule with the surrounding counter-ions form a neutral
entity, the electric double layer, which is of intense experimental and 
theoretical interest \cite{Attard96,Hansen00,Messina09}.
The effective interaction between equivalent electric double layers may lead,
at small enough temperatures, to anomalous attraction of like-charge 
macromolecules \cite{Gulbrand84,Kjellander84}.

The three-dimensional (3D) model of mobile counter-ions near charged
rectilinear hard walls became, due to its relative simplicity, 
the cornerstone for developing theoretical approaches 
to statistical mechanics of classical Coulomb systems.
The high-temperature (weak-coupling) limit is described rigorously by the 
Poisson-Boltzmann (PB) mean-field theory \cite{Andelman} and by its
systematic improvement via the loop expansion 
\cite{Attard88,Netz00,Podgornik90}.
The low-tempe\-ra\-tu\-re (strong-coupling) limit is more controversial
and still under discussion, see \cite{Boroudjerdi05,Netz01,Samaj11a}.
In the leading strong-coupling order, the common feature of all approaches 
is the single-particle picture of counter-ions in the linear 
surface-charge potential.
Recently, the strong-coupling approach based on the Wigner crystallization
\cite{Samaj11a} was generalized to curved cylinder \cite{Mallarino13} and 
disc \cite{Mallarino15} hard walls. 

The standard Coulomb fluids like the one-component or two-component plasmas 
are ``dense'' since the number of mobile charges is proportional to 
the volume of the confining domain. 
Such classical systems exhibit good screening properties.
The bulk particle correlations have a short-range, usually exponential, 
decay at large distances.
The same holds for the decay of particle densities to their bulk values 
in semi-infinite plasmas.
There exists a variety of the exact sum rules in classical Coulomb fluids
which relate the particle one-body and two-body densities, for an old review 
see \cite{Martin88}. 
In the bulk, the charge-charge correlation functions satisfy, 
in any dimension, the zeroth-moment and second-moment Stillinger-Lovett 
screening conditions \cite{Stillinger68a,Stillinger68b}.
In two dimensions, also the compressibility \cite{Baus78,Vieillefosse75} and 
higher-moment \cite{Jancovici00c,Kalinay00} sum rules have been derived.
In the semi-infinite case of an electric double layer, the density of 
particles at the wall was shown to be related to the bulk pressure
via the contact theorem \cite{Choquard80,Mallarino14,Totsuji81}.
The Carnie and Chan generalization of the second-moment Stillinger-Lovett
condition to inhomogeneous fluids leads to the dipole sum rule 
\cite{Carnie81,Carnie83}.
The charge-charge correlation function decays slowly as an inverse-power law
along the wall \cite{Jancovici82a,Usenko76} and the amplitude function
satisfies a sum rule \cite{Jancovici82b,Jancovici95,Samaj10}.
A relation between this algebraic tail and the dipole moment was
established in Ref. \cite{Jancovici01}. 
For finite systems, the information analogous to the bulk second-moment
Stillinger-Lovett rule is contained in the dielectric susceptibility
tensor which, according to phenomenological electrostatics, depends on the 
shape of the confining domain, even in the thermodynamic limit \cite{Landau}.
The microscopic explanation of this shape dependence, due to a contribution
from the long-ranged charge-charge correlations along the domain boundary, 
was given in a series of papers \cite{Choquard86,Choquard87,Choquard89} 
by Choquard et al.

Our system of counter-ions near walls with charged surfaces is ``sparse''
in the sense that, due to the overall electric charge neutrality, the number 
of mobile charges is proportional to the charged surface boundary of 
the domain to which the counter-ions are confined.
The bulk density of particles goes to zero as the volume of the domain
increases.
The screening properties of counter-ions are not good.
Like for instance, in the PB limit the density profile of counter-ions
near one charged plane goes to zero with a slow inverse-power-law decay of 
type $1/x^2$ at large distances $x\to\infty$ from the wall \cite{Andelman}.
Since the sum rules were derived under the assumption of good screening
properties of Coulomb fluids, the poor screening in systems of counter-ions 
near charged surfaces calls into question their validity. 

Although the dimension three is of primary physical interest, it is
useful to go to the two-dimensional (2D) Euclidean space.
In an infinite $d$-dimensional Euclidean space, the electrostatic 
potential $v$ at a point ${\bf r}\in {\rm R}^{d}$, induced by a unit charge 
at the origin ${\bf 0}$, is the solution of the Poisson equation
\begin{equation} \label{Poisson}
\Delta v({\bf r}) = - s_d \delta({\bf r}) ,
\end{equation}
where $s_d$ is the surface area of the unit sphere in $d$ dimensions:
$s_2=2\pi$, $s_3=4\pi$, etc.
This definition of the $d$-dimensional Coulomb potential maintains
generic properties, such as screening sum rules, of ``real'' 3D Coulomb
system with $v({\bf r})=1/r$, $r=\vert {\bf r}\vert$.
In two dimensions, the solution of (\ref{Poisson}), subject to the boundary 
condition $\nabla v({\bf r})\to 0$ as $r\to\infty$, reads 
$v({\bf r}) = - \ln(r/L)$ where the scale $L$ is free.
For counter-ions of charge $-e$ at the inverse temperature 
$\beta=1/(k_{\rm B}T)$, the relevant coupling constant $\Gamma=\beta e^2$.
The main point in two dimensions is that one can solve exactly the model, 
besides the PB mean-field limit $\Gamma\to 0$, also at a specific coupling 
$\Gamma=2$ at which the model is mappable onto free fermions 
\cite{Jancovici81,Jancovici92}.
For specific geometries with known ground-state Wigner structures, one can 
even solve the strong-coupling limit $\Gamma\to\infty$ and make expansions
around this limit.
Moreover, for $\Gamma$ an even positive integer, there exists 
a representation of the partition function and many-body densities 
in terms of anticommuting Grassmann variables \cite{Samaj95,Samaj00,Samaj04}.
The 2D system of counter-ions between symmetrically and asymmetrically 
charged lines for the cylinder geometry was solved in Refs. \cite{Samaj11b} 
and \cite{Samaj14}, respectively.
The validity of certain sum rules for the case of single charged line
was verified in Ref. \cite{Samaj13}. 

The aims of this paper are more ambitious.
We consider a finite 2D disc geometry of the counter-ions model to go to 
the semi-infinite geometry of one charged line which is of primary interest.
The free energy of the particles inside the disc is found in the PB limit 
$\Gamma\to 0$ and at the free-fermion point $\Gamma=2$; the finite-size 
expansion of the free energy does not contain universal term as is usual 
for Coulomb fluids.
Within the Grassmann representation, we derive exact sum rules which are 
related to specific transformations of anticommuting variables. 
These sum rules provide the exact contact density of counter-ions at 
the wall for the disc geometry.  
The amplitude function, which characterizes the asymptotic inverse-power 
law behavior of the two-body density along the wall, is found to be related 
to the particle density profile.
Some of the results obtained in the Poisson-Boltzmann limit are generalized 
to an arbitrary Euclidean dimension.
Performing the M\"obius conformal transformation of particle coordinates
in the partition function, the dielectric susceptibility tensor is 
calculated exactly for an arbitrary coupling and particle number.
It tends to the anticipated disc value in the thermodynamic limit, 
in spite of zero contribution from the bulk region.

The paper is organized as follows.
The basic formalism and studied topics are recapitulated in Sect. \ref{Sect.2}.
Sect. \ref{Sect.3} reviews the formalism of anticommuting Grassmann 
variables to treat the partition function and the many-body densities 
for counter-ions in the disc domain.  
Exactly solvable cases of the free-fermion $\Gamma=2$ and PB $\Gamma\to 0$
coupling constants for the disc geometry are presented in Sect. \ref{Sect.4}.
The sum rules which follow from specific transformations of anticommuting
variables, keeping a composite form of the Grassmann action, 
are summarized in Sect. \ref{Sect.5}.
Here, we derive the contact density of counter-ions at the wall
and the relation between the amplitude function (which characterizes 
the asymptotic inverse-power law behavior of the two-body density along 
the wall) and the particle density profile.
In Sect. \ref{Sect.6}, we study the effect of the M\"obius conformal
transformation of particle coordinates inside the disc on the partition 
function and consequently derive the dielectric susceptibility tensor 
for an arbitrary coupling and particle number.
A brief summary of the obtained results, open problems and conclusions 
are drawn in Sect. \ref{Sect.7}.

\renewcommand{\theequation}{2.\arabic{equation}}
\setcounter{equation}{0}

\section{Formalism and Topics of Interest} \label{Sect.2}

\subsection{Studied geometries}
\begin{figure}[thb] 
\begin{center}
\includegraphics[clip,width=0.85\textwidth]{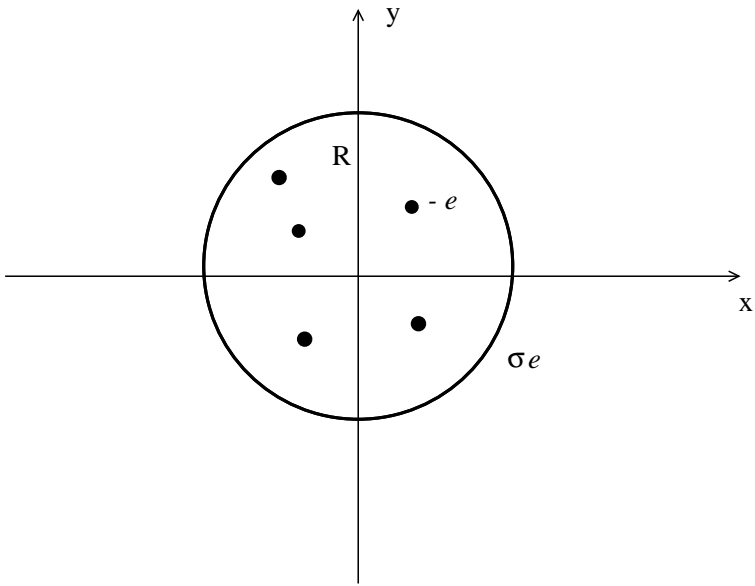} 
\end{center}
\caption{The disc geometry. 
Particles of charge $-e$ are confined to disc of radius $R$. 
There is the uniform line charge density $\sigma e$ on the disc boundary.
The dielectric constants $\epsilon_W$ of the wall $r>R$ and $\epsilon$ of 
the medium in which the particles are immersed are the same, 
$\epsilon_W=\epsilon=1$ in Gauss units.}
\label{Fig:1} 
\end{figure}

The basic 2D geometry we consider is the disc domain of radius $R$, 
$D=\{ {\bf r}, \vert {\bf r}\vert \le R \}$, presented in Fig. \ref{Fig:1}. 
For simplicity, the dielectric constant $\epsilon_W$ of the wall
$W=\{ {\bf r}, \vert {\bf r}\vert > R \}$ is equal to the dielectric
constant $\epsilon$ of medium in which the particles are immersed,
say $\epsilon_W=\epsilon=1$ (vacuum in Gauss units); there are no 
electrostatic image charges.
A constant line charge density $\sigma e$ is fixed on the disc 
circumference $r=R$. 
There are 
\begin{equation} \label{neutr}
N=2\pi R\sigma 
\end{equation}
classical counter-ions of charge $-e$ inside the disc, 
so the system as a whole is electroneutral.
The particle coordinates in the 2D Euclidean space ${\bf r}=(x,y)$ 
can be expressed via radial components $r\in [0,R]$ and $\varphi\in [0,2\pi)$, 
$x=r\cos\varphi$ and $y=r\sin\varphi$, or via the complex variables
$z=r\exp({\rm i}\varphi)$ and $\bar{z}=r\exp(-{\rm i}\varphi)$.
The 2D Coulomb potential is $-\ln(r/L)$ where $L$ is a length scale which
fixes the zero potential at distance $r=L$.

Let $E({\bf r}_1,\ldots,{\bf r}_N)$ be the total Coulomb energy of 
the system for a given configuration of charges at points 
$\{ {\bf r}_i \}_{i=1}^N$.
The partition function at the inverse temperature $\beta=1/(k_{\rm B}T)$ 
is then given by
\begin{equation}
Z_N = \frac{1}{N!} \int_{D} \prod_{i=1}^N {\rm d}^2r_i\,
\exp\left[ -\beta E({\bf r}_1,\ldots,{\bf r}_N) \right] . 
\end{equation} 
The one-body density of particles at point ${\bf r}\in D$ is defined by
\begin{equation}
n({\bf r}) = \langle \hat{n}({\bf r}) \rangle , \qquad
\hat{n}({\bf r}) = \sum_{i=1}^N \delta({\bf r}-{\bf r}_i) ,
\end{equation}
where
\begin{equation}
\langle \cdots \rangle = \frac{\int_{D} \prod_{i=1}^N {\rm d}^2r_i\,
{\rm e}^{-\beta E({\bf r}_1,\ldots,{\bf r}_N)} \cdots}{\int_{D} \prod_{i=1}^N 
{\rm d}^2r_i\,{\rm e}^{-\beta E({\bf r}_1,\ldots,{\bf r}_N)}}
\end{equation}
denotes the statistical average over the canonical ensemble.
Due to the circular symmetry of the problem, we have $n({\bf r})\equiv n(r)$
where $r=\vert {\bf r}\vert$. 
At two-particle level, one introduces the two-body densities
\begin{equation}
n_2({\bf r},{\bf r'}) = \left\langle \sum_{(i\ne j)=1}^N 
\delta({\bf r}-{\bf r}_i) \delta({\bf r}'-{\bf r}_j) \right\rangle .  
\end{equation}
The circular symmetry of the system implies that
$n({\bf r},{\bf r'}) \equiv n(r,r';\varphi-\varphi')$. 
The corresponding (truncated) Ursell functions are defined by
\begin{equation} \label{Ursell}
U({\bf r},{\bf r}') = n_2({\bf r},{\bf r'}) - n({\bf r}) n({\bf r}') . 
\end{equation}
It is useful to introduce also the charge-charge structure function
\begin{eqnarray}
S({\bf r},{\bf r}') & = & e^2 \left[ 
\langle \hat{n}({\bf r}) \hat{n}({\bf r}') \rangle 
- n({\bf r}) n({\bf r}') \right] \nonumber \\ & = &  
e^2 \left[ U({\bf r},{\bf r}') + n({\bf r}) \delta({\bf r}-{\bf r}') \right] . 
\end{eqnarray}

\begin{figure}[thb] 
\begin{center}
\includegraphics[clip,width=0.55\textwidth]{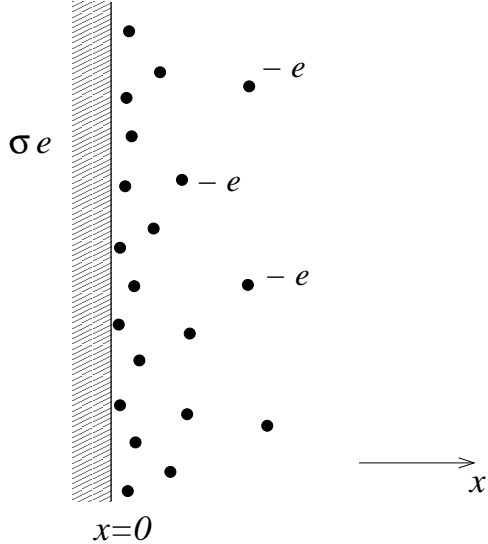} 
\end{center}
\caption{The planar geometry. 
The particles of charge $-e$ are confined to the half-space $x>0$. 
The surface of the wall at $x=0$ carries the uniform surface charge 
density $\sigma e$.}
\label{Fig:2} 
\end{figure}

In $d$ spatial dimensions, we define the semi-infinite planar geometry
in Fig. \ref{Fig:2}.
There is a hard wall in the half-space $x<0$ filled with a material of
dielectric constant $\epsilon_W$, the particles are constrained to 
the half-space $x>0$.
As before, there are no image charges, i.e. $\epsilon_W = \epsilon = 1$. 
Let us represent any vector as ${\bf r}=(x,{\bf y})$, where ${\bf y}$ 
is the set of $(d-1)$ coordinates normal to $x$.
We introduce the same statistical quantities as in the disc case.
Due to invariance with respect to translations along the wall and rotations
around the $x$-axis, we have $n({\bf r})\equiv n(x)$,
$n({\bf r},{\bf r'}) \equiv n(x,x';\vert {\bf y}-{\bf y}'\vert)$, etc. 

For $d=2$, we can pass from the disc to the semi-infinite planar geometry
by taking the inertia to the disc boundary, $r=R-x$, and identifying
\begin{equation} \label{y}
y = 2 R \sin\left( \frac{\varphi}{2} \right) .
\end{equation}
For fixed $x$ and $y$, taking the limits $R\to\infty$ and $\varphi\to 0$
we recover the planar case.
Note that if the angle $\varphi$ is kept finite, the $R\to\infty$ limit 
means simultaneously the asymptotic limit $y\to\infty$.

\subsection{Finite-size corrections of the 2D free energy}
2D critical systems with {\em short-range} interactions among
constituents exhibit universal finite-size properties which are well
understood within the principle of conformal invariance \cite{Cardy}.
For a finite 2D critical system of characteristic size $R$, the 
dimensionless free energy has a large-$R$ expansion of the form
\begin{equation} \label{largeR}
\beta F = A R^2 + B R - \frac{c \chi}{6} \ln R + \cdots .
\end{equation}
The coefficients $A$ and $B$ of the bulk and surface parts, respectively, 
are non-universal.
The coefficient of the term $\ln R$ is universal, dependent only
on the conformal anomaly number $c$ of the critical theory and 
on the Euler number $\chi$ of the manifold ($\chi=1$ for a disc). 
Using plausible arguments, the universal $\ln R$ term was derived 
for dense 2D Coulomb systems at an arbitrary temperature of 
the fluid phase \cite{Forrester91,Jancovici94,Jancovici96}.
Later, the prefactor to the universal $\ln R$ correction was related
to the second-moment of the short-range part of the density-density
direct correlation function $c_{\rm SR}$ \cite{Jancovici00a,Jancovici00b}.
This second moment was evaluated explicitly by using a renormalized
Mayer expansion, for both one-component and two-component plasmas
\cite{Jancovici00c,Kalinay00}.
The obtained results agreed with the prediction of conformal invariance
(\ref{largeR}) for a critical system as we had $c=-1$.
In particular, for the disc of radius $R$ the large-$R$ expansion 
of the free energy is expected to be
\begin{equation} \label{CoulomblargeR}
\beta F = A R^2 + B R + \frac{1}{6} \ln R + \cdots .
\end{equation}

\subsection{Asymptotic decay of correlations along the wall}
The bulk charge correlations decay faster than any inverse power law
in a dense Coulomb fluid as a consequence of perfect screening. 
The screening cloud around a particle sitting near a hard wall is asymmetric.
Consequently, the two-body Ursell functions decay slowly along the wall
for dimensions $d\ge 2$ \cite{Jancovici82a,Usenko76}.
For the plane semi-infinite geometry, using linear response 
\cite{Jancovici82b,Jancovici95,Samaj10} one anticipates an asymptotic 
inverse-power law behavior of type
\begin{equation} \label{asymp}
U(x,x';y) \simeq \frac{f(x,x')}{y^d} , \qquad y\to\infty . 
\end{equation} 
Let us define the $(d-1)$-dimensional Fourier ${\bf k}$-space
with respect to coordinates ${\bf y}$ via
\begin{eqnarray}
U(x,x';y) & = & \int \frac{{\rm d}{\bf k}}{(2\pi)^{d-1}}
\exp({\rm i}{\bf k}\cdot {\bf y}) \hat{U}(x,x';k) , \nonumber \\
\hat{U}(x,x';k) & = & \int {\rm d} {\bf y}
\exp(-{\rm i}{\bf k}\cdot {\bf y}) U(x,x';y) .
\end{eqnarray}
In the sense of distributions \cite{Gelfand}, the asymptotic behavior 
(\ref{asymp}) is governed by the kink at ${\bf k}={\bf 0}$ 
of the small wave number behavior
\begin{equation}
\hat{U}(x,x';k) = \hat{U}(x,x';0) - \frac{s_d}{2} f(x,x') k + \cdots . 
\end{equation}

The function $f(x,x')$ is symmetric in coordinates $x$ and $x'$.
For all exactly solvable cases, it takes the product form
\begin{equation} \label{productform}
f(x,x') = - g(x) g(x') .
\end{equation}
In other words, when the lateral distance between the points ${\bf r}$
and ${\bf r}'$ goes to infinity the $x$-coordinates of the two points
become uncorrelated.
This fact might have intuitively a more general validity. 
The function $f(x,x')$ obeys the sum rule 
\cite{Jancovici82b,Jancovici95,Samaj10} 
\begin{equation} \label{fx}
\int_0^{\infty} {\rm d}x \int_0^{\infty} {\rm d}x'\, f(x,x')
= - \frac{2}{\beta e^2 s_d^2} .
\end{equation}

The asymptotic behavior (\ref{asymp}), formulated for the semi-infinite
planar geometry, has its analogue also for finite domains, in the limit
when the size of the domain tends to infinity.
In particular, it was shown for the 2D disc geometry \cite{Jancovici02}
that, as the radius $R\to\infty$, the Ursell function of two particles
at finite distances $x$ and $x'$ from the boundary and with the angle
$\varphi$ between them behaves as      
\begin{equation} \label{asympdisc}
U(x,x';\varphi) \mathop{\sim}_{R\to\infty} \frac{f(x,x')}{[2R\sin(\varphi/2)]^2} .
\end{equation} 
Here, $2R\sin(\varphi/2)$ is the counterpart of the lateral distance $y$
for the planar geometry, see Eq. (\ref{y}).

\subsection{Dielectric susceptibility tensor}
For dense Coulomb fluids, the small-$k$ behavior of the Fourier transform
of the Coulomb potential gives rise to exact constraints for the charge
structure function $S$ \cite{Martin88}.
In the bulk, $S({\bf r},{\bf r}') = S(\vert {\bf r}-{\bf r}'\vert)$ obeys 
the Stillinger-Lovett screening rules \cite{Stillinger68a,Stillinger68b},
namely the zeroth-moment (electroneutrality condition
\begin{equation} \label{zeroth}
\int {\rm d}^d r, S({\bf r}) = 0 
\end{equation}
and the second-moment condition
\begin{equation} \label{second}
\beta \int {\rm d}^d r\, \vert {\bf r}\vert^2 S({\bf r}) = 
- \frac{2d}{s_d} . 
\end{equation}

For dense Coulomb systems in a finite domain $D$, the analog of 
the bulk zeroth-moment sum rule
\begin{equation} \label{fzeroth}
\int_D {\rm d}^d r'\, S({\bf r},{\bf r'}) = 
\int_D {\rm d}^d r\, S({\bf r},{\bf r'}) = 0 
\end{equation}
holds only in the canonical ensemble with the fixed total charge in $D$.
The information analogous to the bulk second-moment sum rule is contained
in the dielectric susceptibility tensor $\chi_D$ relating the average 
polarization to a constant applied electric field, in the linear limit. 
Representing formally the vector as ${\bf r}=(r^1,r^2,\ldots,r^d)$,
the components of the dielectric susceptibility tensor
$\chi_D^{ij}$ $(i,j=1,\ldots,d)$ are defined by
\begin{equation}
\chi_D^{ij} = \frac{\beta}{\vert D\vert} \int_D {\rm d}^{d}r_1
\int_D {\rm d}^{d}r_2\, r_1^i r_2^j S({\bf r}_1,{\bf r}_2) .
\end{equation}
In the canonical ensemble where the sum rule (\ref{fzeroth}) applies,
this representation is equivalent to the one
\begin{equation}
\chi_D^{ij} = -\frac{\beta}{2 \vert D\vert} \int_D {\rm d}^{d}r_1
\int_D {\rm d}^{d}r_2\, (r_1^i-r_2^j)^2 S({\bf r}_1,{\bf r}_2) .
\end{equation}
As $D\to {\rm R}^{d}$ one might naively anticipate that only the diagonal
components $\chi^i=\lim_{D\to {\rm R}^{d}} \chi_D^{ii}$ $(i=1,\ldots,d)$
are nonzero and that they tend, based on the bulk second-moment 
sum rule (\ref{second}), to the uniform Stillinger-Lovett (SL) value
\begin{equation} \label{SL}
\chi_{\rm SL}^i = - \frac{\beta}{2} \int {\rm d}^{d}r\, (r^i)^2 S({\bf r})
= \frac{1}{s_{d}}
\end{equation}
which does not depend on the shape of domain $D$.
This result is correct for boundary-free domains like the surface of a sphere.
But for the domain with a boundary, it was shown by Choquard et al 
\cite{Choquard86,Choquard87,Choquard89} that the dielectric 
susceptibility is made up of a bulk contribution, which saturates quickly 
to the SL value (\ref{SL}), and of a surface contribution. 
The surface contribution does not vanish in the thermodynamic limit due
to the long-range inverse-power-law decay of particle correlations along 
the boundary (\ref{asymp}).
Summing up both contributions one gets the dielectric susceptibility tensor
whose components {\em depend} on the shape of $D$.
This phenomenon is in agreement with macroscopic electrostatics for
homogeneously polarizable systems \cite{Choquard86,Choquard87,Landau}.
In particular, for the isotropic sphere in the $d$-dimensional 
Euclidean space, defined by the constraint of coordinates
$\sum_{i=1}^{d} (r^i)^2 = R^2$, it holds
\begin{equation} \label{sphere}
\chi_D^i = \frac{d}{s_{d}} , \qquad i=1,\ldots,d .
\end{equation}
This result was verified on the 2D disc geometry, in the high-temperature
Debye-H\"uckel limit and at the exactly solvable coupling $\Gamma=2$
of the OCP.
The extension of the proof to $\Gamma$ an even positive integer
was done in Ref. \cite{Samaj00}.
The generalization to all kinds of dense Coulomb fluids, based on the fact
that the total force acting on a system in thermal equilibrium is zero,
was made in Ref. \cite{Jancovici04}.  

All that has been said holds for dense Coulomb fluids.
For sparse Coulomb systems with charged walls only, the bulk regime 
is trivial with no particles and so there are no bulk sum rules at 
one's disposal. 
The above scenario of the shape-dependent dielectric susceptibility 
tensor is therefore questionable.
 
\renewcommand{\theequation}{3.\arabic{equation}}
\setcounter{equation}{0}

\section{Mapping the Model to 1D Fermions for Disc Geometry} \label{Sect.3}
For the disc geometry presented in Fig. \ref{Fig:1}, the potential induced by 
the line charge density $\sigma e$ is constant inside the disc domain $D$ 
and, with regard to the electroneutrality condition (\ref{neutr}), reads as 
$- N e \ln(R/L)$.
For a given configuration $\{ {\bf r}_i \}_{i=1}^N$ of point charges $-e$
inside the disc, the total energy of the system is composed of three parts:
$E({\bf r}_1,\ldots,{\bf r}_N) = E_{ss} + E_{sp} + E_{pp}$.
The interaction of the fixed line-charge with itself is given by
\begin{equation}
E_{ss} = - \frac{1}{2} (Ne)^2 \ln\left( \frac{R}{L} \right) .
\end{equation} 
The interaction of the line-charge with $N$ mobile particles of charge $-e$
is found to be
\begin{equation}
E_{sp} = (Ne)^2 \ln\left( \frac{R}{L} \right) .
\end{equation}   
Finally, the particle-particle interaction part of the energy is the sum 
over all pair Coulomb interactions:
\begin{equation} \label{Epp}
E_{pp} = -e^2 \sum_{(i<j)=1}^N \ln \frac{\vert z_i-z_j \vert}{L} .
\end{equation}
Introducing the coupling constant $\Gamma\equiv 2\gamma=\beta e^2$,
the corresponding Boltzmann factor at the inverse temperature 
$\beta$ reads as
\begin{equation}
{\rm e}^{-\beta E({\bf r}_1,\ldots,{\bf r}_N)} = 
A_N \prod_{(i<j)=1}^N \vert z_i-z_j\vert^{2\gamma} ,
\qquad A_N = \frac{L^{\gamma N}}{R^{\gamma N^2}} .
\end{equation}
The partition function is given by
\begin{equation} \label{part}
Z_N(\gamma) = \frac{A_N}{\lambda^{2N}} \tilde{Z}_N(\gamma) , \qquad
\tilde{Z}_N(\gamma) = \int_D \prod_{i=1}^N {\rm d}^2 r_i\, 
\prod_{(i<j)=1}^N \vert z_i-z_j\vert^{2\gamma} ,
\end{equation}
where $\lambda$ is the de Broglie wavelength. 
The prefactor $A_N$ is important when computing the free energy
$F_N(\gamma)$ defined by $\beta F_N(\gamma) = - \ln Z_N(\gamma)$.
On the other hand, $A_N$ is irrelevant and can be neglected in the calculation
of the one-body, two-body, $\ldots$ densities.

Our model is a member of a large class of 2D Coulomb systems with the
partition function of the form
\begin{equation} \label{partgen}
\tilde{Z}_N(\gamma) = \int_D \prod_{i=1}^N 
\left[ {\rm d}^2 r_i\, w({\bf r}_i) \right] 
\prod_{(i<j)=1}^N \vert z_i-z_j\vert^{2\gamma} ,
\end{equation}
where $D$ is the confining domain (disc in our case) and $w({\bf r})$ is
the one-body Boltzmann factor which involves all external potentials acting 
on particles. 
Note that the partition function (\ref{part}) corresponds to vanishing
external potential, $w({\bf r})=1$.
The one-body and two-body densities can be obtained in the standard way:
\begin{eqnarray}
n({\bf r}) & = & w({\bf r}) \frac{1}{\tilde{Z}_N} 
\frac{\delta \tilde{Z}_N}{\delta w({\bf r})},
\\ n_2({\bf r},{\bf r}') & = & w({\bf r})  w({\bf r}') \frac{1}{\tilde{Z}_N} 
\frac{\delta^2 \tilde{Z}_N}{\delta w({\bf r}) \delta w({\bf r}')} .
\end{eqnarray}

It has been shown in Ref. \cite{Samaj95} that the partition function
of the form (\ref{partgen}) with the circularly symmetric 
$w({\bf r})\equiv w(r)$ can be expressed in terms of anticommuting
variables $\{ \xi_i^{(\alpha)},\psi_i^{(\alpha)} \}$ $(\alpha=1,\ldots,\gamma)$,
defined on a discrete chain of $N$ sites $i=0,1,\ldots,N-1$, as follows
\begin{equation} \label{antipart}
\tilde{Z}_N(\gamma) = \int {\cal D}\psi {\cal D}\xi\, {\rm e}^{S(\xi,\psi)} , 
\qquad S(\xi,\psi) = \sum_{i=0}^{\gamma(N-1)} \Xi_i w_i \Psi_i .
\end{equation}
Here, ${\cal D}\psi {\cal D}\xi \equiv \prod_{i=0}^{N-1} {\rm d}\psi_i^{(\gamma)}
\cdots {\rm d}\psi_i^{(1)} {\rm d}\xi_i^{(\gamma)} \cdots {\rm d}\xi_i^{(1)}$
and the action $S(\xi,\psi)$ involves pair interactions of composite
operators
\begin{equation} \label{composite}
\Xi_i = \sum_{i_1,\ldots,i_{\gamma}=0\atop (i_1+\cdots+i_{\gamma}=i)}^{N-1}
\xi_{i_1}^{(1)} \cdots \xi_{i_{\gamma}}^{(\gamma)} , \qquad
\Psi_i = \sum_{i_1,\ldots,i_{\gamma}=0\atop (i_1+\cdots+i_{\gamma}=i)}^{N-1}
\psi_{i_1}^{(1)} \cdots \psi_{i_{\gamma}}^{(\gamma)} ,
\end{equation} 
i.e. the products of $\gamma$ anticommuting variables with the fixed
sum of site indices.
The interaction strength is given by
\begin{equation} 
w_i = \int_D {\rm d}^2 r\, r^{2i} w(r) , \qquad i=0,1,\ldots,\gamma(N-1) .
\end{equation}
For our counter-ion system with $w(r)=1$, we have
\begin{equation} \label{wi}
w_i = 2\pi \int_0^R {\rm d}r\, r^{2i+1} = \frac{\pi}{i+1} R^{2(i+1)} .
\end{equation}

The formalism of anticommuting variables was developed further and applied 
to various Coulomb systems in 
Refs. \cite{Samaj00,Samaj04,Samaj11b,Samaj13,Samaj14}.
The main advantage is that the one-body and two-body densities are expressible
explicitly in terms of averaging over the anticommuting variables,
$\langle \cdots\rangle \equiv \int {\cal D}\psi {\cal D}\xi\, 
{\rm e}^S \cdots/\tilde{Z}_N(\gamma)$.
Namely,
\begin{eqnarray}
n(r) & = & w(r) \sum_{i=0}^{\gamma(N-1)} \langle \Xi_i \Psi_i \rangle
r^{2i} , \label{antione} \\
n_2({\bf r}_1,{\bf r}_2) & = & w(r_1) w(r_2) 
\sum_{i_1,j_1,i_2,j_2=0\atop (i_1+i_2=j_1+j_2)}^{\gamma(N-1)} 
\langle \Xi_{i_1} \Psi_{j_1} \Xi_{i_2} \Psi_{j_2} \rangle
z_1^{i_1} \bar{z}_1^{j_1} z_2^{i_2} \bar{z}_2^{j_2} . \label{antitwo}
\end{eqnarray}

\renewcommand{\theequation}{4.\arabic{equation}}
\setcounter{equation}{0}

\section{Exactly Solvable Cases for the 2D Disc Geometry} \label{Sect.4}

\subsection{Free-fermion $\Gamma=2$ coupling}
At $\Gamma=2$ $(\gamma=1)$, the composite operators (\ref{composite})
become the ordinary anticommuting variables.
The partition function (\ref{antipart}) with the diagonal action
$S(\xi,\psi) = \sum_{i=0}^{N-1} \xi_i w_i \psi_i$ is equal to the product
\begin{equation}
\tilde{Z}_N = \prod_{i=0}^{N-1} w_i .
\end{equation}
The two-correlators in the representation of the one-body density 
(\ref{antione}) are given by
\begin{equation}
\langle \xi_i\psi_i \rangle \equiv \frac{1}{\tilde{Z}_N}
\int {\cal D}\psi {\cal D}\xi {\rm e}^{S(\xi,\psi)} \xi_i \psi_i
= \frac{\partial}{\partial w_i} \ln \tilde{Z}_N = \frac{1}{w_i} .
\end{equation}
The four-correlators in the representation of the two-body density 
(\ref{antitwo}) can be calculated using the Wick theorem, with the result
\begin{equation}
\langle \xi_i \psi_j \xi_{i'} \psi_{j'} \rangle = \frac{1}{w_i w_{i'}}
\left( \delta_{ij} \delta_{i'j'} - \delta_{ij'} \delta_{i'j} \right) .
\end{equation}
The first product of Kronecker symbols $\delta_{ij} \delta_{i'j'}$
leads to the term $n(r_1) n(r_2)$ which should be subtracted from
$n_2({\bf r}_1,{\bf r}_2)$ to obtain the corresponding Ursell function.
We recall that for our one-body weight $w(r)=1$, the moments $w_i$
are given by (\ref{wi}). 

The partition function (\ref{part}) reads as
\begin{equation}
Z_N(1) = \frac{L^N}{R^{N^2}\lambda^{2N}} \prod_{j=0}^{N-1} 
\frac{\pi R^{2(j+1)}}{j+1} = 
\left( \frac{\pi L R}{\lambda^2} \right)^N \frac{1}{N!} .
\end{equation}
Using Stirling's asymptotic formula
$\ln N! = N \ln N - N + \frac{1}{2}\ln(2\pi N) +O(1/N)$,
we get for the (dimensionless) free energy
\begin{equation} \label{Ffree}
\beta F_N = - R (2\pi\sigma) \ln\left( \frac{e L}{2\sigma \lambda^2} \right) 
+ \frac{1}{2} \ln (\sigma R) + \ln(2\pi) + O\left( \frac{1}{R} \right) .
\end{equation}
Comparing to the large-$R$ expansion for the free energy of dense Coulomb
fluids (\ref{CoulomblargeR}), $A=0$ because the number of particles
$\propto R$ and the logarithmic term has the prefactor $1/2$ different
from the universal one $1/6$. 

The particle density profile is given by the series
\begin{equation} \label{series}
n(r) = \frac{1}{\pi R^2} \sum_{j=0}^{N-1} (j+1) \left(\frac{r}{R}\right)^{2j} .
\end{equation}
At the boundary, we have
\begin{equation} \label{boundary}
n(R) = 2\pi\sigma^2 + \frac{\sigma}{R} .
\end{equation}
In the planar limit $R\to\infty$, the contact theorem (for a zero 
background charge density) \cite{Choquard80,Mallarino14,Totsuji81}
tells us that the contact particle density equals to 
$\pi\Gamma\sigma^2$ which is in agreement with our $\Gamma=2$ result 
(\ref{boundary}). 
The term $\sigma/R$ is the finite-size correction due to curvature of
the disc boundary.
For $r<R$, the series representation (\ref{series}) implies
\begin{equation}
n(r) = \frac{1}{\pi R^2} \left[
\frac{1-\left(\frac{r}{R}\right)^{2N}}{\left( 1-\frac{r^2}{R^2} \right)^2}
- N \frac{\left(\frac{r}{R}\right)^{2N}}{1-\frac{r^2}{R^2}} \right] .
\end{equation}
In the planar limit $R\to\infty$, using the distance from the wall $x=R-r$
as the variable, we get the profile
\begin{equation}
n(x) = \frac{1}{4\pi x^2} \left[ 1 - (1+4\pi\sigma x) 
{\rm e}^{-4\pi\sigma x} \right] .
\end{equation}
A $x\to\infty$, the density goes to zero with a long-range tail,
$n(x)\simeq 1/(4\pi x^2)$.

The two-body Ursell function admits the series representation
\begin{eqnarray}
U({\bf r}_1,{\bf r}_2) & = & - \frac{1}{(\pi R^2)^2}
\sum_{i=0}^{N-1} (i+1) \left( \frac{z_1\bar{z}_2}{R^2} \right)^i
\sum_{j=0}^{N-1} (j+1) \left( \frac{\bar{z}_1 z_2}{R^2} \right)^j \nonumber \\
& = & - n\left( r_1 r_2 {\rm e}^{{\rm i}\varphi/2}\right)
n\left( r_1 r_2 {\rm e}^{-{\rm i}\varphi/2}\right) , \label{Urs}
\end{eqnarray}
where we introduced the angle difference $\varphi\equiv \varphi_1-\varphi_2$.
Inserting the explicit representations
\begin{equation}
n\left( r_1 r_2 {\rm e}^{\pm{\rm i}\varphi/2}\right) = \frac{1}{\pi R^2} 
\left[ \frac{1-\left(\frac{r_1 r_2}{R^2}\right)^N {\rm e}^{\pm{\rm i}N\varphi}}{
\left( 1-\frac{r_1r_2}{R^2} {\rm e}^{\pm{\rm i}\varphi} \right)^2}
- N \frac{\left(\frac{r_1r_2}{R^2}\right)^N {\rm e}^{\pm{\rm i}N\varphi}}{
1-\frac{r_1r_2}{R^2}{\rm e}^{\pm{\rm i}\varphi}} \right]
\end{equation}
into (\ref{Urs}) and considering the limit $R\to\infty$,
it can be shown that the most relevant contribution comes from 
the product of the terms proportional to $N$:  
\begin{equation}
U(r_1,r_2;\varphi) \mathop{\simeq}_{R\to\infty} - \frac{N^2}{(\pi R^2)^2}
\frac{\left( \frac{r_1 r_2}{R^2} \right)^{2N}}{
1 - 2\cos\varphi \left( \frac{r_1 r_2}{R^2} \right) + 
\left( \frac{r_1 r_2}{R^2} \right)^2} . 
\end{equation}
Going to coordinates $x=R-r$ and taking the planar limit $R\to\infty$,
we finally get the asymptotic behavior of type (\ref{asympdisc}) with
\begin{equation}
f(x,x') = - 4\sigma^2 {\rm e}^{-4\pi\sigma x} {\rm e}^{-4\pi\sigma x'} .
\end{equation} 

To calculate the isotropic dielectric susceptibility tensor $\chi_D^i$
for the disc $D$, we use the identity 
$(x_1x_2+y_1y_2)/2 = (z_1\bar{z}_2+\bar{z}_1z_2)/4$, so that
\begin{equation}
\chi_D^i = \frac{\beta}{\pi R^2} \int_D {\rm d}^2 r_1 \int_D {\rm d}^2 r_2\,
\frac{z_1\bar{z}_2+\bar{z}_1z_2}{4} S({\bf r}_1,{\bf r}_2) .
\end{equation}
Since the structure function is expressible as
\begin{equation}
\frac{S({\bf r}_1,{\bf r}_2)}{e^2} = 
- \sum_{i_1,j_1,i_2,j_2=0\atop (i_1+i_2=j_1+j_2)}^{N-1}
\frac{1}{w_{i_1}w_{i_2}} \delta_{i_1j_2} \delta_{i_2j_1} z_1^{i_1} \bar{z}_1^{j_1}
z_2^{i_2} \bar{z}_2^{j_2} + n(r_1) \delta({\bf r}_1-{\bf r}_2) ,
\end{equation}
we find that 
\begin{equation} \label{chifree}
\chi_D^i = \frac{\Gamma}{2\pi R^2} \left( - \sum_{i=0}^{N-2} \frac{w_{i+1}}{w_i}
+ \sum_{i=0}^{N-1} \frac{w_{i+1}}{w_i} \right) = 
\frac{1}{\pi R^2} \frac{w_N}{w_{N-1}} = \frac{1}{\pi} \frac{N}{N+1} .
\end{equation}
In the $N\to\infty$ limit, we get $\chi_D^i=1/\pi$ which reproduces correctly
the $d=2$ case of the general result (\ref{sphere}).

\subsection{Poisson-Boltzmann theory}
For a given density profile $n({\bf r})$ of counter-ions with charge $-e$,
the the electrostatic potential $\psi({\bf r})$ satisfies the Poisson equation 
\begin{equation}
\Delta \psi({\bf r}) = 2\pi e n({\bf r}) .
\end{equation}
Within the PB mean-field theory, the counter-ions density is locally
related to the corresponding Boltzmann factor of the electrostatic potential
as follows
\begin{equation}
n({\bf r}) = n_0 \exp[\beta e\psi({\bf r})] ,
\end{equation}
where $n_0$ is the normalization factor.
We introduce the reduced potential $\phi({\bf r})\equiv \beta e \psi({\bf r})$ 
for which the above relations read as
\begin{equation} \label{PB}
\Delta \phi({\bf r}) = 2\pi \Gamma n({\bf r}) , \qquad
n({\bf r}) = n_0 {\rm e}^{\phi({\bf r})} .
\end{equation}
For the 2D case of circularly symmetric disc, the counter-ion density and
the electrostatic potential depend only on $r=\vert {\bf r}\vert$ and 
the Laplacian 
\begin{equation}
\Delta = \frac{1}{r} \frac{{\rm d}}{{\rm d}r} \left( r 
\frac{{\rm d}}{{\rm d}r} \right) = \frac{{\rm d}^2}{{\rm d}r^2}
+ \frac{1}{r} \frac{{\rm d}}{{\rm d}r} .
\end{equation}
Integrating the PB equation (\ref{PB}) over the disc and knowing that
the total particle number is $N$, we get the boundary conditions
\begin{equation} \label{BC}
\lim_{r\to 0} r \frac{{\rm d}\phi}{{\rm d}r} = 0 , \qquad
R \frac{{\rm d}\phi}{{\rm d}r}\Big\vert_{r=R} = \Gamma N .
\end{equation}

The general solution of the PB equation (\ref{PB}) takes the form
\begin{equation}
\phi(r) = - 2 \ln (a^2-r^2) + \ln(8 a^2) - \ln(2\pi\Gamma n_0) .
\end{equation}
The boundary condition at $r=R$ (\ref{BC}) fixes the parameter $a$ to
\begin{equation}
a^2 = R^2 + \frac{4 R^2}{\Gamma N} .
\end{equation}
The corresponding density profile at $r\le R$ behaves as
\begin{equation} \label{nr}
n(r) = \frac{4 a^2}{\pi\Gamma} \frac{1}{(a^2-r^2)^2} .
\end{equation}
In particular, the contact density of counter-ions at the wall is given by
\begin{equation} \label{contactPB}
n(R) = \pi\Gamma \sigma^2 + \frac{2\sigma}{R}.
\end{equation}
In the planar limit $R\to\infty$, going to the $x=R-r$ variable,
the density profile reads
\begin{equation}
n(x) = \frac{b\sigma}{(x+b)^2} , \qquad b = \frac{1}{\pi\Gamma\sigma} .
\end{equation}
This is the special 2D case of a more general PB result \cite{Samaj13} 
\begin{equation} \label{nxnu}
n(x) = \frac{b\sigma}{(x+b)^2} , \qquad b = \frac{2}{\beta e^2\sigma s_{d}}
\end{equation}
valid for an arbitrary $d$-dimensional Euclidean space.

As concerns the PB free energy, the particle-particle interaction 
energy (\ref{Epp}) should be calculated via the mean-field
prescription \cite{Hansen00} 
\begin{equation}
E_{pp} = \frac{e^2}{2} \int_0^R {\rm d}^2 r \int_0^R {\rm d}^2 r'\,
n(r) \left( - \ln\vert {\bf r}-{\bf r}'\vert \right) n(r') ,
\end{equation}
where $n(r)$ is the PB profile (\ref{nr}).
Using the formula (see, e.g. \cite{Jancovici01})
\begin{equation}
-\ln\vert {\bf r}-{\bf r}'\vert = - \ln r_> + \sum_{l=1}^{\infty}
\frac{1}{l} \left( \frac{r_<}{r_>}\right)^l \cos[l(\varphi-\varphi')]
\end{equation}
with $r_<=\min(r,r')$ and $r_>=\max(r,r')$, after lengthy algebra
$E_{pp}$ is given by
\begin{equation}
-\beta E_{pp} = \frac{4}{\Gamma} \left[ 
\ln\left( 1 + \frac{\Gamma N}{4} \right) - \frac{\Gamma N}{4}
+ \frac{(\Gamma N)^2}{8} \ln R \right] .
\end{equation} 
Performing the large-$N$ expansion for a fixed $\Gamma$ and substituting
$N\propto R$, we end up with the free energy
\begin{equation} \label{PBfree}
\beta F_N = B R - \frac{4}{\Gamma} \ln (\sigma R) + O(1) .
\end{equation}
This confirms that the $\ln R$ finite-size term is non-universal for
our model of counter-ions.
Note that the $\ln R$ term in the $\Gamma=2$ free energy (\ref{Ffree})
has even the opposite sign. 

It is probably impossible to solve explicitly the two-body density problem
for our finite disc with the density profile (\ref{nr}). 
For the plane geometry in spatial dimension $d$, the two-body density 
was derived in Ref. \cite{Samaj13} and the function $f(x,x')$ has 
the long-range form
\begin{equation}
f(x,x') = - \frac{8}{\beta e^2 s_{d}^2} \frac{b^4}{(x+b)^3(x'+b)^3} ,
\end{equation}
where the parameter $b$ is defined analogously as for the density 
profile in (\ref{nxnu}).

\renewcommand{\theequation}{5.\arabic{equation}}
\setcounter{equation}{0}

\section{Sum Rules for an Arbitrary Coupling} \label{Sect.5}
There exist specific transformations of anticommuting $\xi$-variables
which keep the composite nature of the $\Xi$-operators (\ref{composite}).
Such transformations were applied to the case of the 2D one-component plasma
in Ref. \cite{Samaj00}.
They lead to certain trivial and non-trivial constraints among correlators
of the composite operators which are the same for all one-component systems.
However, these constraints lead to sum rules for one-body and two-body
densities whose forms {\em depends} on the particular model. 
Although all sum rules are derived strictly for $\gamma$ a positive integer,
one can assume their extended validity for all real values of $\gamma$
in the fluid regime. 

\subsection{Sum rules for one-body density}
First we rescale by constant $\mu$ one of the field components, say
\begin{equation} \label{trans1}
\xi_i^{(1)} \to \mu \xi_i^{(1)} , \qquad i=0,1,\ldots,N-1 .
\end{equation}
The composite $\Xi$-operators transform simply as $\Xi_i\to \mu\Xi_i$.
The invariance of the partition function $Z_N$ (\ref{antipart}) with respect
to this transformation around $\mu=1$ leads to 
(see \cite{Samaj00} with the diagonalized weights $w_{ij}=w_i\delta_{ij}$)
\begin{equation} \label{con1}
\sum_{i=0} w_i \langle \Xi_i\Psi_i \rangle = N .
\end{equation}
With respect to the representation (\ref{antione}) of the one-body density,
this constraint provides a trivial sum rule
\begin{equation}
\int_D {\rm d}^2 r\, n(r) = N .
\end{equation}
It tells us that the number of particles in domain $D$ equals to $N$,
which is the well-known information.
This sum rule holds for all one-component system, independently of
the one-body Boltzmann weight $w(r)$.

Let us now consider another transformation for all $\xi$-field
components
\begin{equation} \label{trans2}
\xi_i^{(\alpha)} \to \lambda^i \xi_i^{(\alpha)} , \qquad i=0,1,\ldots,N-1 ;
\quad \alpha=1,\ldots,\gamma .
\end{equation}
The composite $\Xi$-operators transform as $\Xi_i\to \lambda^i \Xi_i$.
The invariance of the partition function (\ref{antipart}) with respect
to this transformation leads to \cite{Samaj00}
\begin{equation} \label{con2}
\sum_{i=0} i w_i \langle \Xi_i\Psi_i \rangle = \frac{1}{2} \gamma N(N-1) .
\end{equation}
Summing this nontrivial constraint with the previous one (\ref{con1})
and recalling that for our counter-ion system with $w(r)=1$ the circular 
moments are given by (\ref{wi}), we obtain
\begin{equation}
\sum_{i=0}^{\gamma(N-1)} \langle \Xi_i \Psi_i \rangle \pi R^{2(i+1)}
= \frac{1}{2} \gamma N(N-1) + N .
\end{equation}
This relation is equivalent to the exact formula for the contact density 
of counter-ions at the wall
\begin{equation} \label{contactnR}
n(R) = \pi \Gamma \sigma^2 + \left( 2-\frac{\Gamma}{2} \right) 
\frac{\sigma}{R} .
\end{equation}
In the planar limit $R\to\infty$, the contact particle density is equal to
$\pi\Gamma\sigma^2$ in agreement with the contact theorem
\cite{Choquard80,Mallarino14,Totsuji81}.  
The $1/R$ correction due to the wall curvature is exact for an arbitrary
coupling $\Gamma$. 
For the free-fermion coupling $\Gamma=2$ and in the PB limit $\Gamma\to 0$, 
we reproduce the exact results (\ref{boundary}) and (\ref{contactPB}),
respectively.

\subsection{Sum rules for two-body densities}
The transformation of the anticommuting variables (\ref{trans1}), when 
applied to $Z_N \langle\Xi_i\Psi_i\rangle$, leads to \cite{Samaj00}
\begin{equation}
\sum_{j=0}^{\gamma(N-1)} w_j \langle \Xi_i \Psi_i \Xi_j \Psi_j \rangle
= (N-1) \langle \Xi_i \Psi_i \rangle . 
\end{equation}
For an arbitrary one-component system, it can be easily shown that this 
constraint is equivalent to the sum rule
\begin{equation} \label{sr2}
n({\bf r}) = - \int_{D} {\rm d}{\bf r}'\, U({\bf r},{\bf r}') 
\end{equation}
which is equivalent to the zeroth-moment sum rule (\ref{fzeroth}).

The transformation (\ref{trans2}), when applied to 
$Z_N \langle\Xi_i\Psi_i\rangle$, leads to \cite{Samaj00}
\begin{equation} \label{sr3}
\sum_{j=0}^{\gamma(N-1)} j w_j \langle \Xi_i \Psi_i \Xi_j \Psi_j \rangle
= \left[ \frac{1}{2} \gamma N(N-1) - i \right] \langle \Xi_i \Psi_i \rangle . 
\end{equation}
For $w(r)=1$, the fermionic representation of the two-body density 
(\ref{antitwo}) can be written in polar coordinates as follows
\begin{equation}
n_2({\bf r}_1,{\bf r}_2) = \sum_{i_1,j_1,i_2,j_2=0\atop (i_1+i_2=j_1+j_2)}^{\gamma(N-1)} 
\langle \Xi_{i_1} \Psi_{j_1} \Xi_{i_2} \Psi_{j_2} \rangle
r_1^{i_1+j_1} {\rm e}^{{\rm i}\varphi_1(i_1-j_1)} r_2^{i_2+j_2} 
{\rm e}^{{\rm i}\varphi_2(i_2-j_2)} .
\end{equation}
We perform the operation 
\begin{eqnarray}
\frac{r_2}{2} \frac{\partial}{\partial r_2} n_2({\bf r}_1,{\bf r}_2) 
& = & \sum_{i_1,j_1,i_2,j_2=0\atop (i_1+i_2=j_1+j_2)}^{\gamma(N-1)} 
\langle \Xi_{i_1} \Psi_{j_1} \Xi_{i_2} \Psi_{j_2} \rangle
r_1^{i_1+j_1} {\rm e}^{{\rm i}\varphi_1(i_1-j_1)} \nonumber \\ & &
\times \frac{i_2+j_2}{2} r_2^{i_2+j_2} {\rm e}^{{\rm i}\varphi_2(i_2-j_2)} 
\end{eqnarray}
and then integrate over ${\bf r}_2$ inside the disc.
The integration over the angle $\varphi_2$ fixes $i_2=j_2=j$ and, 
consequently, $i_1=j_1=i$.
Thus we get
\begin{eqnarray}
\int_0^R {\rm d}^2 r_2\, \frac{r_2}{2} 
\frac{\partial}{\partial r_2}  n_2({\bf r}_1,{\bf r}_2) & = &
\sum_{i,j=0}^{\gamma(N-1)} \langle \Xi_i\Psi_i \Xi_j\Psi_j \rangle r_1^{2i} 
j w_j \nonumber \\ & = &
\sum_{i=0}^{\gamma(N-1)} \left[ \frac{1}{2}\gamma N(N-1) - i \right]
\langle \Xi_i\Psi_i \rangle r_1^{2i} , 
\end{eqnarray}  
where we applied the sum rule (\ref{sr3}).
Using formula (\ref{con2}), this relation can be readily rexpressed 
in terms of the Ursell function (\ref{Ursell}) as follows
\begin{equation}
\int_0^{2\pi} {\rm d}\varphi \int_0^R {\rm d}r_2\, r_2^2
\frac{\partial}{\partial r_2} U(r_1,r_2;\varphi) = 
- r_1 \frac{\partial n(r_1)}{\partial r_1} .  
\end{equation}
Integrating by parts and using (\ref{sr2}), we finally arrive at
\begin{equation} \label{sumrule1}
R^2 \int_0^{2\pi} {\rm d}\varphi\, U(r,R;\varphi)
= - r \frac{\partial n(r)}{\partial r} - 2 n(r) .
\end{equation}

Writing $r=R-x$, $y=R\varphi$ and going to the planar limit $R\to\infty$, 
we obtain the 2D version of the WLMB (Wertheim, Lovett, Mou, Buff) equation
\cite{Lovett76,Wertheim76} 
\begin{equation}
\int_{-\infty}^{\infty} {\rm d}y\, U(0,x;y) = \frac{\partial n(x)}{\partial x} .
\end{equation}

In Sect. 4 of Ref. \cite{Samaj00} it was shown that the linear
transformation of all anticommuting variables
\begin{equation}
\xi_i^{(\alpha)}(t) = \sum_{j=i}^{N-1} {j\choose i} t^{j-i} \xi_j^{(\alpha)} ,
\qquad i=0,\ldots,N-1; \quad \alpha=1,\ldots,\gamma  
\end{equation}  
implies the ``compact'' transformation of composite variables
\begin{equation}
\Xi_i(t) = \sum_{j=i}^{\gamma(N-1)} {j\choose i} t^{j-i} \Xi_j ,
\qquad i=0,\ldots,\gamma(N-1) .
\end{equation} 
The invariance of $Z_N\langle \Xi_i\Psi_i\rangle$ with respect to
the parameter $t$, considered around $t=0$, implies the constraint
\begin{equation} \label{sr4}
\sum_{j=0}^{\gamma(N-1)-1} w_j (j+1) 
\langle \Xi_i \Psi_{i+1} \Xi_{j+1} \Psi_j \rangle =
- (i+1) \langle \Xi_{i+1} \Psi_{i+1} \rangle .
\end{equation}
To make use of this relation, we consider the integral
$\int_0^{2\pi}{\rm d}\varphi_2\, z_1 \bar{z}_2 n_2({\bf r}_1,{\bf r}_2)$.
Inserting the fermionic representation of the two-body density
(\ref{antitwo}) with $w(r)=1$ into this integral, only terms with 
$j_2=i_2-1\equiv j$ and $i_1=j_1-1\equiv i$ survive and we obtain
\begin{equation}
\int_0^{2\pi}{\rm d}\varphi_2\, z_1 \bar{z}_2 n_2({\bf r}_1,{\bf r}_2)
= 2\pi \sum_{i,j=0}^{\gamma(N-1)-1} 
\langle \Xi_i \Psi_{i+1} \Xi_{j+1} \Psi_j \rangle
(z_1 \bar{z}_1)^{i+1} (z_2 \bar{z}_2)^{j+1} .
\end{equation}
Now let us fix the point 2 at the boundary, i.e. $z_2\bar{z}_2=R^2$,
and use the complex notation $z_1=r{\rm e}^{{\rm i}\varphi}$.
Considering $w_j(j+1) = \pi R^{2(j+1)}$ in the constraint (\ref{sr4}), 
we find that
\begin{eqnarray}
\int_0^{2\pi}{\rm d}\varphi'\, r R {\rm e}^{{\rm i}(\varphi-\varphi')}
n_2(r,R;\varphi-\varphi') & = & -2 \sum_{i=0}^{\gamma(N-1)-1} (i+1)
\langle \Xi_{i+1}\Psi_{i+1} \rangle r^{2(i+1)} \nonumber \\
& = & -2 \sum_{i=0}^{\gamma(N-1)} i \langle \Xi_i\Psi_i \rangle r^{2i} 
\nonumber \\ & = & - r \frac{\partial n(r)}{\partial r} .
\end{eqnarray}
Finally, fixing $\varphi=0$, using the obvious symmetry 
$n_2(r,R;\varphi') =n_2(r,R;-\varphi')$ and noting that the substitution
of the two-body density by the Ursell function has no effect on 
the integral, we arrive at the sum rule
\begin{equation} \label{sumrule2}
R \int_0^{2\pi} {\rm d}\varphi\, \cos\varphi\, U(r,R;\varphi) =
- \frac{\partial n(r)}{\partial r} .  
\end{equation}

From the exact sum rules (\ref{sumrule1}) and (\ref{sumrule2}),
we can construct a combination which implies
\begin{equation}
\frac{1}{2} \int_0^{2\pi} {\rm d}\varphi\, 
\left[ 2 R \sin\frac{\varphi}{2} \right]^2 U(r,R;\varphi) 
= (R-r) \frac{\partial n(r)}{\partial r} - 2 n(r) .
\end{equation}
In the limit $R\to\infty$, switching to the variable $x=R-r$ and
using the special case of the asymptotic formula (\ref{asympdisc}) 
\begin{equation}
U(x,0;\varphi) \mathop{\sim}_{R\to\infty} \frac{f(x,0)}{[2R\sin(\varphi/2)]^2} ,
\end{equation}
we get the relation
\begin{equation} \label{final}
f(x,0) = - \frac{1}{\pi} \left[ x \frac{\partial n(x)}{\partial x} 
+ 2 n(x) \right] 
\end{equation}
valid for the semi-infinite 2D planar geometry.

It is simple to check by using the findings in Sect. 4 that this 2D relation 
indeed holds for the free-fermion coupling $\Gamma=2$,
\begin{equation}
n(x) = \frac{1}{4\pi x^2} \left[ 1 - (1+4\pi\sigma x) 
{\rm e}^{-4\pi\sigma x} \right] , \qquad
f(x,0) = - 4\sigma^2 {\rm e}^{-4\pi\sigma x}  
\end{equation}
as well as in the PB limit $\Gamma\to 0$, 
\begin{equation}
n(x) = \frac{\sigma b}{(x+b)^2} , \qquad 
f(x,0) = - \frac{1}{\pi} \frac{2\sigma b^2}{(x+b)^3} ,
\end{equation}
where $b=1/(\pi\Gamma\sigma)$.

\subsection{Possible generalization to any dimension}
Let us assume that the function $f(x,x')$ factorizes into the product form 
(\ref{productform}) not only for all exactly solvable cases, but also
for all temperatures.
Under this assumption our 2D result (\ref{final}) generalizes to 
\begin{equation} \label{finalgen}
f(x,x') = - \frac{1}{2\pi^2\Gamma\sigma^2} 
\left[ x \frac{\partial n(x)}{\partial x} + 2 n(x) \right]
\left[ x' \frac{\partial n(x')}{\partial x'} + 2 n(x') \right] ,  
\end{equation}
where the prefactor was fixed on the base of the 2D planar contact theorem
$n(0)=\pi\Gamma\sigma^2$.
The consistency of the formalism is confirmed by the fact that the formula 
(\ref{finalgen}) fulfills the 2D version of the sum rule (\ref{fx})
\begin{equation}
\int_0^{\infty} {\rm d}x \int_0^{\infty} {\rm d}x'\, f(x,x')
= - \frac{1}{2\pi^2\Gamma}
\end{equation} 
owing to the equality
\begin{equation} \label{formula}
\int_0^{\infty} {\rm d}x\, \left[ 
x \frac{\partial n(x)}{\partial x} + 2 n(x) \right]
= \int_0^{\infty} {\rm d}x\, n(x) = \sigma .
\end{equation}
Here, we applied the integration by parts to the term 
$x\partial n(x)/\partial x$ together with the known fact that $n(x)$ goes 
to 0 faster than $1/x$ as $x\to\infty$ and the electroneutrality condition.

For an arbitrary dimension $d$, the counterpart of the formula 
(\ref{finalgen}) reads as 
\begin{equation} \label{finalgennu}
f(x,x') = - \frac{2}{\beta e^2 s_{d}^2\sigma^2} 
\left[ x \frac{\partial n(x)}{\partial x} + 2 n(x) \right]
\left[ x' \frac{\partial n(x')}{\partial x'} + 2 n(x') \right] .  
\end{equation}
In view of the relation (\ref{formula}), the sum rule (\ref{fx}) is satisfied. 
The PB solution presented in Sect. 4, 
\begin{equation}
n(x) = \frac{\sigma b}{(x+b)^2} , \qquad 
f(x,x') = - \frac{8}{\beta e^2 s_{d}^2} \frac{b^4}{(x+b)^3(x'+b)^3}
\end{equation}
where $b=2/(\beta e^2 s_{d}\sigma)$, is consistent with the ansatz
(\ref{finalgennu}).
Whether or not this ansatz holds also beyond the PB limit is an open
question. 

\renewcommand{\theequation}{6.\arabic{equation}}
\setcounter{equation}{0}

\section{Microscopic Calculation of the Dielectric Susceptibility Tensor} 
\label{Sect.6}

\subsection{M\"obius conformal transformation}
We consider the particles with complex coordinates $(z,\bar{z})$
inside the disc domain $D=\{ (z,\bar{z}), z\bar{z} \le R^2 \}$. 
The M\"obius conformal transformation
\begin{equation}
z' = \frac{r_0 z + R^2}{z+\bar{r}_0} , \qquad
z = \frac{\bar{r}_0 z' - R^2}{-z'+ r_0} 
\end{equation}
(with the free complex parameter $r_0$ such that $r_0 \bar{r}_0 \ne R^2$)
transforms the particle coordinates in the disc domain $D$ to another
domain $D'$ defined by the inequality
\begin{equation}
(R^2-r_0\bar{r}_0) ( R^2 - z'\bar{z}') \le 0 .
\end{equation}
If $r_0$ is chosen outside the disc, i.e. 
\begin{equation}
r_0\bar{r}_0>R^2 ,
\end{equation} 
we obtain that $z'\bar{z}'\le R^2$.
This means that the original disc domain $D$ is mapped onto itself, $D'=D$.

Let us study the effect of the M\"obius transformation of all particle
coordinates on the partition function
\begin{eqnarray} 
\tilde{Z}_{\gamma}(N) & = & \frac{1}{N!} \int_D \prod_{i=1}^N {\rm d} z_i {\rm d} 
\bar{z}_i\, W_{\gamma}(z_1,\ldots,z_N) , \nonumber \\
W_{\gamma}(z_1,\ldots,z_N) & = & \prod_{(i<j)=1}^N \vert z_i-z_j\vert^{2\gamma} .
\label{part1}
\end{eqnarray}
Each surface element ${\rm d}z{\rm d}\bar{z}$ transforms as
\begin{equation}
{\rm d}z {\rm d}\bar{z} = \frac{(r_0\bar{r}_0-R^2)^2}{
(r_0-z')^2(\bar{r}_0-\bar{z}')^2} {\rm d}z' {\rm d}\bar{z}' 
\end{equation}
and the square of the distance between two particles transforms as
\begin{equation}
\vert z_i-z_j\vert^2 = \frac{(r_0\bar{r}_0-R^2)^2}{
(r_0-z'_i)(\bar{r}_0-\bar{z}'_i)(r_0-z'_j)(\bar{r}_0-\bar{z}'_j)} 
\vert z'_i-z'_j\vert^2 .
\end{equation}
The partition function (\ref{part1}) can be written in terms of 
the transformed coordinates as follows
\begin{equation} \label{part2}
\tilde{Z}_{\gamma}(N) = \frac{1}{N!} 
\int_D \prod_{i=1}^N {\rm d} z_i {\rm d} \bar{z}_i
\left[ \frac{r_0\bar{r}_0-R^2}{(r_0-z_i)(\bar{r}_0-\bar{z}_i)} 
\right]^{\gamma(N-1)+2} W_{\gamma}(z_1,\ldots,z_N) .
\end{equation}
Here, in order to simplify the notation, we have omitted the prime in all 
particle coordinates.
Up to a trivial prefactor which does not depend on particle coordinates,
the M\"obius transformation induces at the fixed point $(r_0,\bar{r}_0)$
outside the disc $(r_0\bar{r}_0>R^2)$ the opposite charge $-e Q$ with $Q$ 
given by $2\gamma Q = \gamma(N-1)+2$. 

\subsection{Dielectric susceptibility tensor}
The partition functions (\ref{part1}) and (\ref{part2}) are identical.
To take advantage of this identity, we place the induced charge $-e Q$
at infinity, i.e. $r_0\bar{r}_0\to \infty$.
In this limit, each one-body factor in (\ref{part2}) can be expanded
as follows
\begin{eqnarray}
\left[ \frac{r_0\bar{r}_0-R^2}{(r_0-z_i)(\bar{r}_0-\bar{z}_i)} \right]^{\nu}
& = & \left[ \frac{1-\frac{R^2}{r_0\bar{r}_0}}{\left( 1-\frac{z_i}{r_0}\right)
\left( 1-\frac{\bar{z}_i}{\bar{r}_0}\right)} \right]^{\nu} \nonumber \\
& = & 1 + \nu \frac{z_i}{r_0} + \nu \frac{\bar{z}_i}{\bar{r}_0}
- \nu \frac{R^2}{r_0\bar{r}_0} + \nu^2 \frac{z_i\bar{z}_i}{r_0\bar{r}_0}
+ \cdots , 
\end{eqnarray}
where we introduced the notation $\nu\equiv\gamma(N-1)+2$.
To make a systematic expansion in the variables $1/r_0$ and $1/\bar{r}_0$,
we first take 1 from each one-body factor in the representation (\ref{part2}) 
and obtain nothing but the partition function (\ref{part1}).
All other expansion terms must vanish.
The terms proportional to $1/r_0$ or $1/\bar{r}_0$ trivially vanish.
As concerns the next expansion terms proportional to $1/(r_0\bar{r}_0)$,
there are $N$ one-body contributions of type $-\nu R^2 + \nu^2 z_i\bar{z}_i$
and $N(N-1)/2$ two-body contributions of type 
$\nu^2 z_i \bar{z}_j + \nu^2 \bar{z}_i z_j$ 
for each ordered particle pair $(i<j)$.
The consequent constraint reads as
\begin{eqnarray}
\frac{1}{N!} \int_D \prod_{i=1}^N {\rm d} z_i {\rm d} \bar{z}_i\,
W_{\gamma}(z_1,\ldots,z_N) \sum_{i,j=1}^N z_i \bar{z}_j 
\phantom{aaaaaaaa} \nonumber \\ 
= \frac{R^2 N}{\gamma(N-1)+2} \, \frac{1}{N!} \int_D \prod_{i=1}^N {\rm d} z_i 
{\rm d} \bar{z}_i\, W_{\gamma}(z_1,\ldots,z_N) .
\end{eqnarray}
In terms of the statistical averages, this relation is equivalent to
\begin{equation} \label{sum1}
\int_0^R {\rm d}^2 r \int_0^R {\rm d}^2 r'\, {\bf r}\cdot {\bf r}'
\langle \hat{n}({\bf r}) \hat{n}({\bf r}') \rangle 
= \frac{R^2 N}{\gamma(N-1)+2} .
\end{equation}
Note that the same equality holds for the truncated correlator
$\langle \hat{n}({\bf r}) \hat{n}({\bf r}') \rangle - n(r) n(r')$ since 
\begin{equation} \label{sum2}
\int_0^R {\rm d}^2 r \int_0^R {\rm d}^2 r'\, {\bf r}\cdot {\bf r}'
n(r) n(r') = 0
\end{equation}
after the integration of $\cos(\varphi-\varphi')$ over the angle 
$\varphi-\varphi'$ from $0$ to $2\pi$.

In view of the relations (\ref{sum1}) and (\ref{sum2}), the diagonal $i=x,y$ 
elements of the isotropic dielectric susceptibility tensor are given by
\begin{equation} \label{finalformula}
\chi_{\rm disc}^i = \frac{\beta}{\pi R^2} \int_0^R {\rm d}^2 r 
\int_0^R {\rm d}^2 r'\, \frac{{\bf r}\cdot {\bf r}'}{2} S({\bf r},{\bf r}')
= \frac{1}{\pi} \frac{\gamma N}{\gamma(N-1)+2} . 
\end{equation} 
It is interesting that this exact result is available for an arbitrary
coupling $\Gamma=2\gamma$ and an arbitrary number of particles $N$.
For the free-fermion case $\gamma=1$, we reproduce the result (\ref{chifree}).
In the thermodynamic limit $N\to\infty$, we obtain the expected value
$\chi_{\rm disc}^i = 1/\pi$.

\renewcommand{\theequation}{7.\arabic{equation}}
\setcounter{equation}{0}

\section{Conclusion} \label{Sect.7}
The system of counter-ions near walls charged by a fixed surface charge density
has poor screening properties and a trivial bulk regime with no particles.
The Coulomb sum rules are valid for dense Coulomb systems like 
the one-component or two-component plasmas and in their derivation good
screening properties and, in certain cases, the non-trivial ``dense'' bulk
regime are crucial.
Some of these sum rules were confirmed for the counter-ion system 
in Ref. \cite{Samaj13}.

In this paper, we treated the system of counter-ions in the 2D disc of radius 
$R$ and subsequently went to the semi-infinite model by increasing 
$R\to\infty$. 
Using the formalism of anticommuting variables for the partition function
and many-body densities of 2D one-component Coulomb systems, we derived
specific sum rules valid for any coupling constant equal to an even integer.
One of these sum rules fixes the contact density of counter-ions at
the disc wall, see formula (\ref{contactnR}).   
In the semi-infinite 2D planar geometry, we have established the exact relation 
between the amplitude function $f(x,0)$ (which characterizes the asymptotic 
inverse-power law behavior of the two-body density along the wall) 
and the particle density profile in Eq. (\ref{final}).
If one assumes that for any coupling constant the function $f(x,x')$ exhibits 
the product form (\ref{productform}), this function is determined
unambiguously via Eq. (\ref{finalgen}).
The generalization to an arbitrary Euclidean dimension $d$ (\ref{finalgennu}),
respecting the sum rule (\ref{fx}), works well in the PB limit. 
Whether or not the formula (\ref{finalgennu}) is applicable to any dimension
$d$ and to any temperature is an open question.
Another open question is whether an analogous relation between $f(x,x')$
and the density profile holds for dense Coulomb fluid like the one-component
or two-component plasmas.

The finite-$R$ expansion of the free energy for particles being inside 
the disc of radius $R$ was found at the free-fermion point $\Gamma=2$ 
(\ref{Ffree}) and in the PB limit (\ref{PBfree}).
The prefactor to the term $\ln R$, which is universal in the case of 
standard dense Coulomb fluids, depends on $\Gamma$ for our model.

The microscopic calculation of the dielectric susceptibility tensor
in Sect. \ref{Sect.6} is based on the M\"obius conformal transformation
of particle coordinates in the definition of the partition function.
It is interesting that the final formula (\ref{finalformula}) holds for
any coupling and particle number.
Although the bulk contribution to the tensor elements is equal to zero, 
their values are consistent with the prediction of macroscopic electrostatics.
There might exist a counterpart of the M\"obius transformation for 
higher-dimensional spheres, with a potential generalization of 
the present result. 

\begin{acknowledgements}
The support received from Grant VEGA No. 2/0015/15 is acknowledged. 
\end{acknowledgements}

\end{document}